\documentclass[twocolumn,aps,prc,floatfix]{revtex4}
\usepackage{graphicx}
\usepackage{dcolumn}
\usepackage{bm}
\usepackage{mhchem}

\begin{document}

\title{Blind spots of probing the high-density symmetry energy in heavy-ion collisions}

\author{Gao-Chan Yong} 

\affiliation{%
{Institute of Modern Physics, Chinese Academy of Sciences, Lanzhou
730000, China}
}%

\begin{abstract}
The nuclear symmetry energy, especially at suprasaturation densities, plays crucial roles in many astrophysical studies. However, nowadays the high-density behavior of the symmetry energy is still very controversial in nuclear community. To constrain the high-density behavior of the symmetry energy, neutron-rich nuclei collisions at medium energies are considered to be one of the most effective methods. While probing the high-density symmetry energy by using heavy-ion collisions, blind spots may exist.
In the framework of the Isospin-dependent Boltzmann-Uehling-Uhlenbeck (IBUU) transport model,
the blind spots of probing the high-density symmetry energy by the n/p ratio in the central Au+Au reaction at 300 MeV/nucleon are demonstrated. It is found that the nucleon observable neutron to proton ratio n/p in heavy-ion collisions cannot effectively probe the high-density symmetry energy when the high-density symmetry energy is less density-dependent.

\end{abstract}


\maketitle


The equation of state (EoS) of
nuclear matter at density $\rho$ and isospin asymmetry
$\delta$ ($\delta=(\rho_n-\rho_p)/(\rho_n+\rho_p)$) can be
expressed as \cite{li08,bar05}
\begin{equation}
E(\rho ,\delta )=E(\rho ,0)+E_{\text{sym}}(\rho )\delta ^{2}+\mathcal{O}%
(\delta ^{4}),
\end{equation}%
where $E_{\text{sym}}(\rho)$ is the nuclear symmetry energy.
Nowadays the EoS
of isospin symmetric nuclear matter $E(\rho, 0)$ is relatively well
determined \cite{pawl2002} but the EoS of isospin
asymmetric nuclear matter, especially the high-density behavior
of the nuclear symmetry energy, is still very uncertain \cite{Guo14}.
There are plenty of studies showing inconsistent
results on pion production \cite{wolter06,xie13,xiao09,prassa07,feng10,hong2014,Reisdorf07,cozma17} when comparing theoretical simulations to the data from FOPI detector at GSI \cite{Reisdorf07}.
Constraining the high-density behavior of the symmetry energy from ground-based measurements is highly relevant to the physics of neutron stars \cite{Lat01}, such as their stellar radii and moments of inertia, crustal vibration frequencies and neutron star cooling rates \cite{Lat04,Vil04,Ste05}, the gravitational-wave frequency \cite{gwf,gwf2} and the gamma-ray bursts \cite{grb} in neutron star mergers \cite{GWth,GW170817}.
The conditions and characteristics of r-process nucleosynthesis depend on the
amount of ejected material and the thermodynamic conditions and matter
composition of the ejecta thus also on the EoS of neutron-rich matter \cite{rpn1,rpn2,rpn3}.
Experimentally, constraints on the high-density symmetry energy by the measurements of pion and nucleon, triton and $^{3}$He yields ratio in the isotope reaction systems $^{132}\rm {Sn}+^{124}\rm {Sn}$ and $^{108}\rm {Sn}+^{112}\rm {Sn}$ at about 300 MeV/nucleon, are being carried out at RIBF-RIKEN in Japan \cite{sep,shan15}. Probing the high-density symmetry energy with other heavy systems at higher incident beam energies are also being carried out/planned at FOPI/GSI and CSR/Lanzhou \cite{fopi16,csr} and some progress has been made by measuring nucleon and light charged cluster flows \cite{fopi16}.

To constrain the high-density behavior of the symmetry energy by heavy-ion collisions, one usually varies the density dependence of the symmetry energy in transport model simulations and
make comparisons with experimental data. This operation is alright for those symmetry energies which
evidently depend on the density. But for the symmetry energy which has a less dependence on the density, the above method is noneffective. Because according to the chemical equilibrium condition of nuclear matter formed in heavy-ion collisions \cite{chemical2002}, only if the symmetry energy changes with density, the liquid-gas phase transition can occur. That is to say, if the value of the symmetry energy is less density-dependent (since the high-density is very controversial \cite{Guo14}, all the density dependences of the high-density symmetry energy are possible), a freeze-out observable in heavy-ion collisions cannot prove the information of the symmetry energy effectively.

To demonstrate the blind spots of probing the high-density symmetry energy in heavy-ion collisions, based on our updated Isospin-dependent Boltzmann-Uehling-Uhlenbeck (IBUU) transport model, we studied the symmetry-energy-sensitive observable neutron to proton ratio
n/p ratio in the central Au+Au reaction at 300 MeV/nucleon. It is shown that the nucleon observable n/p ratio in heavy-ion collisions cannot effectively probe the high-density symmetry energy when a less density-dependent symmetry energy is employed.


The used Isospin-dependent
Boltzmann-Uehling-Uhlenbeck (IBUU) transport model originates from the IBUU04 model \cite{lyz05}.  The effects of neutron-proton short-range-correlations are appropriately taken into account in the mean-field potential and in the initialization of colliding nuclei, respectively \cite{yong20171,yong20172}. And the transition momentum of the minority is set to be the same as that of the majority in asymmetric matter \cite{yong2018}. The in-medium inelastic baryon-baryon collisions and the in-medium pion transport are also considered \cite{yongm2016,yongp2015}.

In the IBUU transport model, the time
evolution of the single particle phase space distribution function
$f(\vec{r},\vec{p},t)$ is described by
\begin{equation}
\frac{\partial f}{\partial
t}+\nabla_{\vec{p}}E\cdot\nabla_{\vec{r}}f-\nabla_{\vec{r}}E\cdot\nabla_{\vec{p}}f=I_{c}.
\label{IBUU}
\end{equation}
The left-hand side of
Eq.~(\ref{IBUU}) gives the time evolution of the particle phase
space distribution function due to its transport and mean field,
and the right-hand side collision item $I_{c}$ accounts for the
modification of the phase space distribution function due to the two body interactions.
The neutron and proton density distributions in nucleus are given by the Skyrme-Hartree-Fock with Skyrme M$^{\ast}$ force parameters \cite{skyrme86}. In projectile and target nuclei, the proton and neutron momentum distributions with high-momentum tails of $p_{max}$= 2 $p_{F}$ are employed \cite{yong20171,yong20172,sci08,sci14,yongcut2017}.

The isospin- and momentum-dependent single nucleon mean-field
potential reads as \cite{yong20171}
\begin{eqnarray}
U(\rho,\delta,\vec{p},\tau)&=&A_u(x)\frac{\rho_{\tau'}}{\rho_0}+A_l(x)\frac{\rho_{\tau}}{\rho_0}\nonumber\\
& &+B\Big(\frac{\rho}{\rho_0}\Big)^{\sigma}(1-x\delta^2)-8x\tau\frac{B}{\sigma+1}\frac{\rho^{\sigma-1}}{\rho_0^\sigma}\delta\rho_{\tau'}\nonumber\\
& &+\frac{2C_{\tau,\tau}}{\rho_0}\int
d^3\,p'\frac{f_\tau(\vec{r},\vec{p^{'}})}{1+(\vec{p}-\vec{p^{'}})^2/\Lambda^2}\nonumber\\
& &+\frac{2C_{\tau,\tau'}}{\rho_0}\int
d^3\,p'\frac{f_{\tau'}(\vec{r},\vec{p^{'}})}{1+(\vec{p}-\vec{p^{'}})^2/\Lambda^2},
\label{buupotential}
\end{eqnarray}
where $\rho_0$ denotes the saturation density, $\tau, \tau'$=1/2(-1/2) is for neutron (proton).
$\delta=(\rho_n-\rho_p)/(\rho_n+\rho_p)$ is the isospin asymmetry,
and $\rho_n$, $\rho_p$ denote neutron and proton densities,
respectively. The parameter values $A_u(x)$ = 33.037 - 125.34$x$
MeV, $A_l(x)$ = -166.963 + 125.34$x$ MeV, B = 141.96 MeV,
$C_{\tau,\tau}$ = 18.177 MeV, $C_{\tau,\tau'}$ = -178.365 MeV, $\sigma =
1.265$, and $\Lambda = 630.24$ MeV/c.
With these settings, the empirical values of nuclear matter at normal density are reproduced, i.e., the saturation density $\rho_{0}$ = 0.16 fm$^{-3}$, the binding energy $E_{0}$ = -16 MeV, the
incompressibility $K_{0}$ = 230 MeV, the isoscalar effective mass
$m_{s}^{*} = 0.7 m$, the single-particle potential
$U^{0}_{\infty}$ = 75 MeV at infinitely large nucleon momentum at
saturation density in symmetric nuclear matter, the symmetry
energy $E_{\rm sym}(\rho_0) = 30$ MeV \cite{yong20171}.
Note here that at nuclear densities studied here, the average kinetic symmetry energy just accounts for less than 20\% of the total symmetry energy thus the symmetry potential plays a major role in the constitution of the high-density symmetry energy. In Eq.~(\ref{buupotential}), different symmetry energy's stiffness parameters $x$
can be used in different density regions to mimic
different density-dependent symmetry energy.

The isospin-dependent baryon-baryon ($BB$) scattering cross section in medium $\sigma
_{BB}^{medium}$ is reduced compared with their free-space value
$\sigma _{BB}^{free}$ by a factor of
\begin{eqnarray}
R^{BB}_{medium}(\rho,\delta,\vec{p})&\equiv& \sigma
_{BB_{elastic, inelastic}}^{medium}/\sigma
_{BB_{elastic, inelastic}}^{free}\nonumber\\
&=&(\mu _{BB}^{\ast }/\mu _{BB})^{2},
\end{eqnarray}
where $\mu _{BB}$ and $\mu _{BB}^{\ast }$ are the reduced masses
of the colliding baryon pairs in free space and medium,
respectively. The effective mass of baryon in isospin asymmetric nuclear matter
is expressed as
\begin{equation}
\frac{m_{B}^{\ast }}{m_{B}}=1/\Big(1+\frac{m_{B}}{p}\frac{%
dU}{dp}\Big).
\end{equation}
More details on the present used model can be found in Ref.~\cite{yong20171}.

\begin{figure}[th]
\centering
\includegraphics[width=0.55\textwidth]{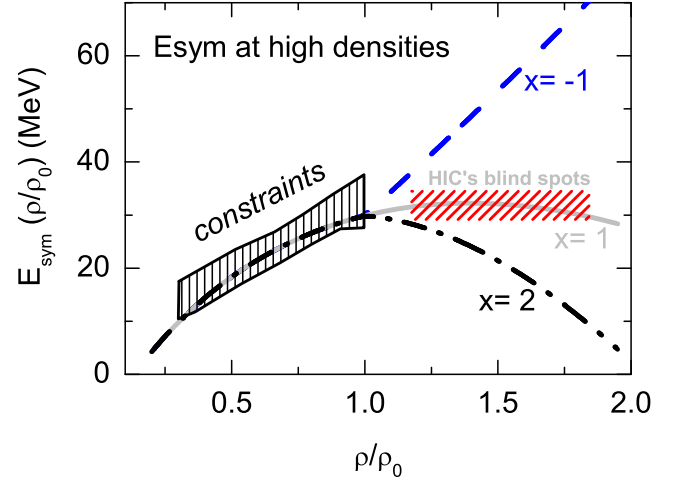}
\caption{The used density-dependent symmetry energy derived from the single particle potential Eq.~(\ref{buupotential}) with different $x$ parameters in different density regions.} \label{esym}
\end{figure}
Fig.~\ref{esym} shows the used symmetry energy derived from the single particle potential Eq.~(\ref{buupotential}) with different $x$ parameters at low and high densities. Because below the saturation density, the symmetry energy is roughly constrained \cite{horowitz2014,lwchen2017}, we fix the form of the low-density symmetry energy with parameter $x$ = 1. From Fig.~\ref{esym}, it is seen that
the low-density symmetry energy with parameter $x$ = 1 is well consistent with the current constraints. Since the symmetry energy at high densities is still not well constrained \cite{Guo14}, in the present study, we vary the high-density symmetry energy parameter $x$ in the range of $x$ = -1, 1, 2. These choices cover the current uncertainties of the high-density behavior of the symmetry energy \cite{Guo14,lwchen2017}. In the study, we also make transport simulations without the high-density symmetry energy.
This is achieved by setting $\rho_{n}= \rho_{p}= (\rho_{n}+ \rho_{p})/2$ and
$U_{n}= U_{p}= (U_{n}+ U_{p})/2$.


\begin{figure}[t]
\centering
\includegraphics[width=0.55\textwidth]{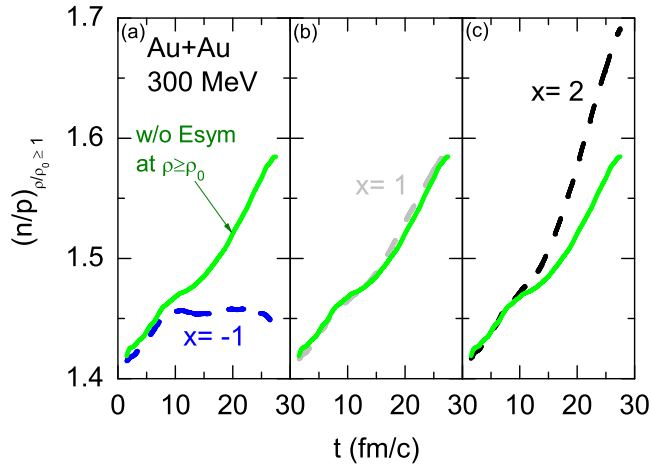}
\caption{Neutron to proton ratio n/p of dense matter ($\rho/\rho_{0}\geq1$) as a function of time in central Au+Au
reaction at 300 MeV/nucleon with different high-density symmetry energies. The solid line denotes the case without high-density symmetry energy.} \label{densernp}
\end{figure}
To probe the high-density symmetry energy in dense matter formed in heavy-ion collisions, it is useful to first see the asymmetry of formed dense matter. Fig.~\ref{densernp} shows the evolution of the neutron to proton ratio n/p of dense matter formed in central Au+Au reaction at 300 MeV/nucleon with different high-density symmetry energies. For each case, we use the same low-density symmetry energy as shown in Fig.~\ref{esym}. In each panel of Fig.~\ref{densernp}, the solid line denotes the case without high-density symmetry energy. From Fig.~\ref{densernp} (a) and (c), compared with the case without high-density symmetry energy, one sees the stiffer high-density symmetry energy ($x$= -1) causes a smaller asymmetry of dense matter while a softer high-density symmetry energy ($x$= 2) causes a larger asymmetry. Both cases are understandable since the high-density symmetry energy is repulsive/attractive for neutrons with the stiff/soft high-density symmetry energy, thus leaving a small/large proportion of neutrons in the dense matter for the the stiff/soft high-density symmetry energy.

While from Fig.~\ref{densernp} (b), compared with
the case without high-density symmetry energy, it is interesting to see that the less density-dependent high-density symmetry energy (with parameter $x$= 1, shown in Fig.~\ref{esym}) almost has no effect on the asymmetry of dense matter formed in heavy-ion collisions. That is to say, the less density-dependent high-density symmetry energy almost does not affect the isospin-fractionation of dense matter formed in heavy-ion collisions \cite{chemical2002}. In fact, for isospin-fractionation of nuclear matter, there is a chemical equilibrium condition \cite{chemical2002,muller,liko,baran,shi}
\begin{equation}\label{chem}
E_{sym}(\rho_1)\delta_1=E_{sym}(\rho_2)\delta_2,
\end{equation}
where $E_{sym}(\rho_1), E_{sym}(\rho_2)$ are the symmetry energies at different density regions and $\delta_1, \delta_2$ are, respectively, the asymmetries of the two different parts of nuclear matter.
$\delta\equiv (\rho_n-\rho_p)/(\rho_n+\rho_p)$ with $\rho_n$ and $\rho_p$ being the neutron and proton densities.
It is energetically
favorable in a dynamical process to have a migration of nucleons
with its direction determined by Eq. (\ref{chem}) according to the
density dependence of the symmetry energy. Since the high-density symmetry energy with parameter $x$= 1 almost does not change its value with the increase of density, i.e., $ E_{sym}(\rho_1) \approx E_{sym}(\rho_2)$, the asymmetry of dense matter would be not affected by such density-dependent symmetry energy, thus in dense matter the isospin-fractionation $\delta_1\approx\delta_2$. Here the asymmetry $n/p$ = $(1+\delta)/(1-\delta)$.
Therefore, compared with the case without high-density symmetry energy, one sees in Fig.~\ref{densernp} (b), the less density-dependent high-density symmetry energy with parameter $x$= 1 almost does not affect the asymmetry of dense matter formed in heavy-ion collisions.
Alternatively, in heavy-ion collisions the less density-dependent high-density symmetry energy may be not observable since it cannot cause isospin fractionation.

\begin{figure}[t]
\centering
\includegraphics[width=0.55\textwidth]{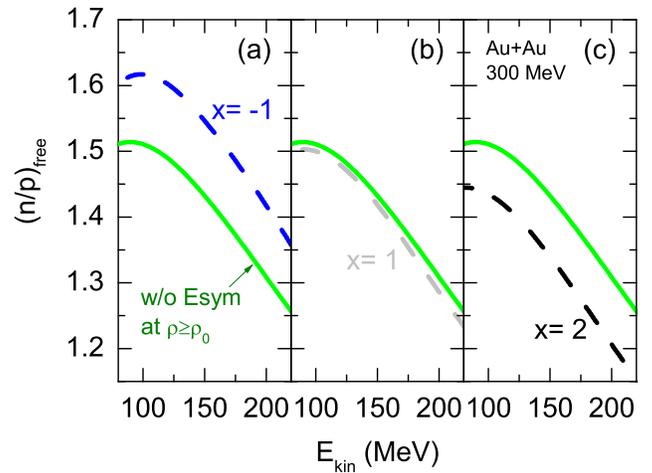}
\caption{Free neutron to proton ratio n/p as a function of kinetic energy in central Au+Au
reaction at 300 MeV/nucleon with different high-density symmetry energies. The solid line denotes the case without high-density symmetry energy.} \label{freernp}
\end{figure}
Because the symmetry potential has opposite signs for neutron and proton
and the fact that the symmetry potential is generally smaller compared to
the isoscalar potential, and also because the symmetry potential acts directly on nucleons and normally nucleon emissions are rather abundant in typical heavy-ion reactions,
the neutron/proton ratio n/p of nucleon emissions may be one of the best observables to probe the symmetry energy \cite{BCY2005,li1997}.

Shown in Fig.~\ref{freernp} is the free neutron to proton ratio n/p as a function of kinetic energy in central Au+Au reaction at 300 MeV/nucleon with different high-density symmetry energies. The free neutron to proton ratio n/p at low kinetic energies is not shown since the low-energy nucleons may be from cluster decays of hot fragments which complicate the question studied here. From Fig.~\ref{freernp} (a), (c), one sees that compared with the case without  high-density symmetry energy the stiffer high-density symmetry energy ($x$= -1) causes a larger free n/p ratio while the softer high-density symmetry energy ($x$= 2) gives a smaller free n/p ratio. This is understandable since the stiffer/softer high-density symmetry energy push more/less neutrons to be free in heavy-ion collisions. As expected, From Fig.~\ref{freernp} (b)
one again sees that the less density-dependent high-density symmetry energy ($x$= 1) does not affect the free n/p ratio evidently. This is because the less density-dependent high-density symmetry energy cannot cause isospin fractionation in dense matter formed in heavy-ion collisions as discussed previously. Therefore, one can conclude that the less density-dependent high-density symmetry energy may be not probed directly in heavy-ion collisions.

One may argue that if the experimental data lies between the results with the stiff ($x$= -1) and the soft ($x$= 2) symmetry energies, the true high-density symmetry energy should also lie between the stiff and the soft symmetry energies. This deduction may be not reliable because the high-density symmetry energy may, for instance, have an abrupt change at certain density point due to the possible chiral phase transition \cite{xuz}. If the symmetry energy keeps less density-dependent in certain density region, its effectiveness in heavy-ion collisions should be roughly the same as that without the symmetry energy.

In nuclear physics community, especially in recent ten years, one usually tries to constrain the high-density behavior of the symmetry energy by rare isotope reactions worldwide, such as at the Facility for Rare Isotope Beams (FRIB) in the Untied States \cite{frib}, the Radioactive Isotope Beam Facility (RIBF) at RIKEN in Japan \cite{sep,shan15}, or the GSI Facility for Antiproton and
Ion Research (FAIR) in Germany \cite{fopi16}, the Cooling Storage Ring on the Heavy Ion Research Facility at IMP (HIRFL-CSR) in China \cite{csr}, and the Rare Isotope Science Project (RISP) in Korea \cite{korea}. However, as we discussed above, one may not directly probe the high-density symmetry energy in heavy-ion collisions in case the high-density symmetry energy is less
density-dependent. From the above studies, it is shown that the effectiveness of the
less density-dependent high-density symmetry energy is roughly equal to the case without high-density symmetry energy. One therefore has to find some other ways to probe the less density-dependent high-density symmetry energy.

The blind spots of probing the high-density symmetry energy in heavy-ion collisions only occur in case the high-density symmetry energy in certain density region is less density-dependent. The blind spots do not appear in the density regions if the high-density symmetry energy is evidently density-dependent.


Based on the isospin-dependent transport model, it is demonstrated that if the high-density symmetry energy is less density-dependent the isospin fractionation could not occur in dense matter. The less density-dependent high-density symmetry energy thus may be not directly probed in rare-isotope collisions. Since the isospin fractionation could not occur in case of the high-density symmetry energy is
less density-dependent, the general symmetry-energy-sensitive observables may not afford to probe the high-density symmetry energy. One should keep this point in mind while constraining the high-density symmetry energy by symmetry-energy-sensitive observables in heavy-ion collisions.
In this case, one way to probe the high-density symmetry energy by heavy-ion collisions is to first measure the ratio of the yields of free neutron and proton in heavy-ion collisions and make comparison with the transport simulation with only the constrained low-density symmetry energy as shown in Fig.~\ref{esym}. If one cannot reproduce experimental data by comparison with the transport simulations, then the high-density symmetry energy is density-dependent, otherwise the high-density symmetry energy is less density-dependent. After confirming the high-density symmetry energy is density-dependent, the symmetry-energy-sensitive observables thus can be used to probe the high-density symmetry energy in heavy-ion collisions.


The author thanks B.A. Li for useful communications. This work is supported in part by the National Natural Science
Foundation of China under Grant Nos. 11775275, 11435014.

\end{document}